\documentclass[aps,prl,reprint,preprintnumbers,nofootinbib]{revtex4-1}
\pdfoutput=1
\usepackage{amsmath}
\usepackage{amssymb}
\usepackage{amsfonts}
\usepackage{hyperref}
\usepackage{epsfig}
\usepackage{graphicx}
\usepackage{xcolor}
\usepackage{color}
\usepackage{slashed} 
\usepackage{mathrsfs} 
\def\beq{\begin{equation}}
\def\eeq{\end{equation}}
\def\bea{\begin{eqnarray}}
\def\eea{\end{eqnarray}}
\def\ba{\begin{array}}
\def\ea{\end{array}}

\def\t1{\tilde{t}_1}
\def\t2{\tilde{t}_2}
\def\b1{\tilde{b}_1}

\newcommand{\bd}{\begin{displaymath}}
\newcommand{\ed}{\end{displaymath}}
\newcommand{\be}{\begin{equation}}
\newcommand{\ee}{\end{equation}}

\def\b{\beta}

\def\q2 {q^2}

\def\t {\times }

\def\bt{\begin{table}}
\def\et{\end{table}}

\catcode`@=11 
\def \gsim{\mathrel{\mathpalette\@versim>}}
\def \lsim{\mathrel{\mathpalette\@versim<}}
\def \@versim#1#2{\lower0.4ex\vbox{\baselineskip\z@skip\lineskip\z@skip
     \lineskiplimit\z@\ialign{$\m@th#1\hfil##\hfil$%
     \crcr#2\crcr\sim\crcr}}}
\catcode`@=12 

\begin{document}

\title{%
Same-sign trileptons as a signal of sneutrino lightest supersymmetric partlcle}%

\author{ Arindam Chatterjee$^1$},
\author{ Nabarun Chakrabarty$^2$, Biswarup Mukhopadhyaya$^2$}%
\affiliation{$^{1}$~ Indian Statistical Institute, 203, B.T. Road, Kolkata- 700108, India} 
\affiliation{$^2$~Regional Centre for Accelerator-based Particle Physics \\
     Harish-Chandra Research Institute\\
Chhatnag Road, Jhusi, Allahabad - 211 019, India}
\begin{abstract}
Contrary to common expectation, a left-sneutrinos can occasionally be
the lightest supersymmetric particle. This has important implications in both
collider and dark matter studies.  We show that same-sign tri-lepton
(SS3L) events at the Large Hadron Collider, with any lepton having opposite sign 
vetoed, distinguish such scenarios, up to gluino masses exceeding 2 TeV.  
The $jets~+~MET$ signal rate is somewhat suppressed in this case, thus 
enhancing the scope of leptonic signals. 

\end{abstract}
 
\maketitle



Supersymmetry (SUSY), or a symmetry between elementary bosons and
fermions, has been a matter of great interest over several decades.
In the form where lepton (L) and baryon (B) numbers are conserved,
SUSY offers a stable particle which is the dark matter (DM) candidate
for the universe. Therefore, physicists not only ponder on possible
discovery channels for SUSY at the Large Hadron Collider (LHC), 
\cite{Mukhopadhyaya:2010qf,Mukhopadhyay:2011xs} but also wish to 
know how, if discovered, we can identify the lightest SUSY particle 
(LSP) which is the DM candidate. In the minimal SUSY standard model 
(MSSM) or its immediate extensions, the DM candidate\cite{Baer:2014eja}
usually is ${\chi^0_1}$, the lightest neutralino (a linear superposition 
of the `partners' of the photon, the Z-boson and the neutral Higgs-like 
spinless particles), the gravitino (partner of the graviton) \cite{Roszkowski:2014lga,Cottin:2014cca}, or the axino (partner of an axion) \cite{Baer:2008eq,Baer:2010kd}. The signals at the LHC are dominantly jets with missing transverse energy (MET) \cite{Chatrchyan:2014lfa,Aad:2014wea} occasionally with leptons and/or photons alongside.


In contrast, it is difficult to have a SUSY spectrum with a left-chiral 
sneutrino (${\tilde \nu}_L$, the spinless partner of a neutrino) as the 
DM candidate. Such an LSP has unsuppressed interaction with the Z-boson 
and is therefore  disfavoured from direct DM search experiments, unless 
its mass is well above a TeV. However, in case this restriction is avoided (as seen below) and one has a (left) sneutrino LSP, finding its distinct signature at the LHC is a desideratum. We show here that the scenario is distinguishable through same-sign trileptons (SS3L) at the LHC. Extensive scans carried out by us \cite{Mukhopadhyaya:2010qf,Mukhopadhyay:2011xs} over the parameter space fail to turn up regions where, in an R-parity conserving SUSY spectrum, containing only superpartners of Standard Model (SM) particles alone, can lead to SS3L signals with such abundance. Moreover, compared to the case of a ${\chi^0_1}$ LSP, the $0 lepton~+~jets~+~MET$ events get suppressed, and the leptonic final states gain more importance, thus warranting a revision of collider search strategies.

A ${\tilde \nu}_L$ DM can be allowed, if there is a mass-splitting between the scalar (${\tilde \nu}_1$) and pseudoscalar (${\tilde
\nu}_2$) components of ${\tilde \nu}_L = {\tilde \nu}_1 + i{\tilde
\nu}_2$ . The Z couples to ${{\tilde \nu}_1}{{\tilde \nu}_2}$. A splitting of a few hundred keV's prevents the scattering of the lighter of ${\tilde \nu}_1$ and ${\tilde \nu}_2$ (which is the DM candidate) into 
the heavier one via such coupling. The energy barrier created by this 
split is insurmountable unless the dark matter candidate has a speed exceeding its escape velocity in our galaxy\cite{Hall:1997ah,TuckerSmith:2001hy,Ma:2011zm,Chatterjee:2014vua}. 
This mass difference can occur, for example, from a tiny Majorana neutrino mass, for which the necessary conditions have been discussed in the literature \cite{Ma:2011zm}. Also, the sneutrino can be the lightest in the MSSM spectrum, just above a gravitino, an axino or even a right-chiral sneutrino LSP. Such spectrum has been considered in \cite{Katz:2009qx,Katz:2010xg}. All these scenarios are addressed by the SS3L signal which is otherwise highly suppressed in  R-parity conserving SUSY where $R~=~(-1)^{(3B+L+2S)}$.

SS3L is inevitable in the scenarios discussed above, because the ${\tilde \nu}_L$ states are close in mass to the charged sleptons (${\tilde l}_L$), as dictated by $SU(2)_L$ invariance. The latter (leaving aside the staus and their mixing) are slightly more massive, mainly because of D-term contributions. Therefore, if the lightest (gaugino-like) neutralino is the next massive state in the spectrum, it decays either to a charged slepton and an anti-lepton (or to its conjugate state) or to the left-sneutrino(s) and a neutrino, with comparable branching ratios. ${\tilde l}_L$ undergoes three-body decays, producing the corresponding sneutrino and two soft-jets or a soft lepton and a neutrino. The soft leptons do not mostly survive the event selection criteria. Thus all SUSY cascades resulting in the lightest neutralino lead to two leptons in about half of the cases. The Majorana nature of neutralinos causes these two leptons to be of the same type in half the cases among such events. Further, a third lepton of the same sign can come from 
cascades, via either a top quark or a chargino. Thus one has three (or even four) leptons of the same sign.\footnote{This leaves out the situation where the lighter chargino is decoupled and the lighter stop is so close to to ${\chi^0_1}$ that it decays only into $c {\chi^0_1}$.} 

Unlike ref.\cite{Katz:2009qx,Katz:2010xg}, our main focus is on SS3L events. Further, contrary to the brief discussion in \cite{Katz:2010xg}, we demonstrate that SS3L may be obtained from a simple spectrum and its observation need not imply the presence of a right-slepton (in addition to a left-slepton doublet) in the low energy spectrum. We emphasize that a simple spectrum with left-sneutrino LSP, without any additional SUSY particles, may lead to the rather
distinct SS3L signal. We also demonstrate that the decay mode $\tilde{t}_1 \rightarrow b \chi^+_1$ adversely affects SS3L events when $\chi_1^{\pm}$ decays into sleptons or sneutrinos.
  
Throughout our discussion we will assume the first two generations of
$SU(2)_L$ doublet sleptons to be degenerate. Further, both $e, ~\mu$
will be described as {\it leptons} ($\ell$), and their scalar counterparts, as {\it sleptons} ($\tilde{\ell}$). Further, since various mechanisms may be responsible for the production of DM in the early Universe \cite{Moroi:1999zb,Acharya:2008bk,Acharya:2009zt} and there may even be additional DM candidate(s) possibly from hidden sector, we will not restrict the collider analysis by assuming thermal production of sneutrinos.

For simplicity, we assume the first two families of squarks to be
decoupled. A stop well within the reach of the LHC is retained, thus
providing a semblance of naturalness, and the gluino is assumed to be
heavier than the stop. Other than the light charged sleptons,
sneutrinos and ${\chi^0_1}$, we have used benchmark points in the SUSY
parameter space with both light and heavy ${\chi^{\pm}_1}$ and
${\chi^0_2}$.  The parameter $\mu$ and thus the Higgsino-dominated
states are kept above a TeV without any loss of generality. The
channels of our interest are both $\tilde{t}_1\tilde{t}_1^*$
production, and cascade production of the lighter stop (or the
anti-stop with the same rate) from the decay of the gluino
($\tilde{g}$).  This is a conservative choice from the viewpoint of
the SS3L signal, since larger event rates should be expected if the
first two families of squarks are also produced.

We assume a bino-like ${\chi^0_1}$ and wino--like ${\chi^{\pm}_1}$ and
${\chi^0_2}$.  When one has a sneutrino LSP, the first two families of
$SU(2)_L$ doublet sleptons are the next-to-lightest ones (assumed to
be degenerate for simplicity).  The stau mass is taken to be at least
a TeV; staus lighter than ${\chi^0_1}$ can cause some reduction to our
predicted signals, but keeps it within the same order of magnitude.
Based on the nature of the intermediate neutralino(s), the following
scenarios have been considered as representative.

\begin{figure}[ht!]
\begin{center}
\includegraphics[scale=0.50,angle=-0]{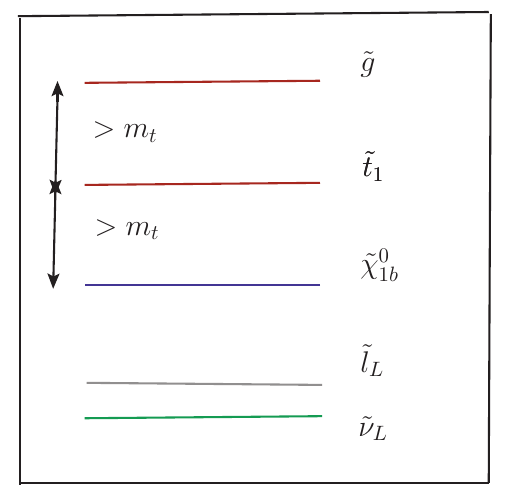}
\end{center}
\caption{The mass hierarchy required to obtain SS3L. In the 
simplest scenario, only a bino–like ${\chi^0_1}$ has been introduced 
     between $\tilde{t}_1$ and the (first two generations of) slepton doublets. }
\label{fig:spec}
\end{figure}

\begin{enumerate}
 \item In the simple scenario (A) with just the $\chi_1^0$
   within reach, direct production of a stop-antistop pair causes each
   (anti)stop to decay directly into ${\chi^0_1}$. While these
   ${\chi^0_1}$'s give rise to two same-sign leptons as already
   explained, the third lepton of the same sign comes from the decay
   of a (anti) top produced in (anti)stop decay. The number of SS3L
   events is further enhanced in the non-decoupling gluino case where
   additional (anti-)stops are produced from $\tilde{g}$ decay. 
   It should be noted that SS4L is also possible, though with a 
   reduced rate, if a pair of gluinos decay into two top-stop pairs. 
   This happens when both the $W$'s, produced from the decay of two      (anti)top quarks, yield leptons of identical sign.
 
 \item In scenarios (B) and (C), in addition to the bino--like
   ${\chi^0_1}$, a wino--like chargino ${\chi^{\pm}_1}$ and the
   corresponding neutralino ${\chi^0_2}$ also occur below $\tilde{t}_1$ in the spectrum.  There is consequently an additional
   decay mode, namely, $\tilde{t}_1 \rightarrow b \chi_1^+$. 
   However, the branching ratio in this channel depends on the  composition of $\tilde{t}_1$.  While $\tilde{t}_1$ is dominantly 
   right-type in scenario (B), significant amount of left-right 
   mixing is allowed in scenario (C).  Because of its large
   hypercharge, an R-type ($SU(2)_L$ singlet) $\tilde{t}_1$ will
   dominantly decay into ${\chi^0_1}$, while for an L-type ($SU(2)_L$
   doublet) $\tilde{t}_1$ there is a substantial branching ratio into
   the $b \chi_1^\pm$ channel. In such a situation, both of the 
   stops in the two decay chains will tend to produce charginos which
   tend to undergo two-body decays into charged sleptons. This makes it
   difficult to have SS3L final states in the direct stop
   pair-production, and one has to depend only on cascades from
   gluino decay. Thus, while both the scenarios B and C include
   a light chargino,  scenario C represents a situation where
   the composition of the lighter stop tends to reduce the rate of SS3L.
   As we shall see below, one still expects to see this signal with a rate
   sufficient to discern the sneutrino-LSP scenario. It should be mentioned
   in addition that both scenarios B and C retain the possibility of
   seeing SS4L (albeit with smaller rates)  whenever SS3L is allowed.
\end{enumerate}
All the three scenarios are allowed by the 8 TeV data so far\cite{Aad:2014pda}. 
The above discussion shows that, while BP A has no wino-like state
affecting the phenomenology, even the presence of such states affects
the suggested SS3L signal only if the lighter stop has a substantial
left component, and thus BP B and BP C have different LHC
implications. While a stop decays almost  entirely into a top and the
${\chi^0_1}$ in BP A, this branching ratio becomes 91\% in BP B and
56\% in BP C.  The branching ratio for $b \chi_1^{\pm}$ ( $t
\chi^0_2$), on the other hand, is 6\% (3\%) and 31\% (13\%),
respectively, for BP B and C. The important aspects of the spectrum
with each of the three benchmark points (BP) mentioned above are
summarized in Table \ref{tab:bp}.  The nature of the spectrum for BP A 
is also shown in Figure 1. Note that the presence of a
right-slepton above the neutralino(s) does not affect the signal.

\begin{table}[ht!]
\begin{center}
\begin{tabular}{|c|c|c|c| } \hline
Parameter	& BP-A	&BP-B  & BP-C\\
\hline
\hline
$m_{\tilde{g}}$           & 1600   & 1600   &  1600 \\
$m_{\tilde{t}_1}$         & 1000   & 1000   &  1000\\
$m_{\chi_1^0}$    & 590    & 441    &  443 \\    
$m_{\chi_2^0}$    & --    &  620    &  620 \\    
$m_{\chi_1^+}$    & --    &  620    &  620 \\    
$m_{\tilde{\nu}}$         & 293    & 293    &  293 \\    
\hline
\hline
\end{tabular}
\end{center}
\caption{ Mass spectra for different benchmark points. BP-A and BP-B represent 
scenario (A) (with only the bino-like neutralino intermediate state) and scenario (B) 
(with a bino-like and a wino-like neutralino together with a wino-like chargino 
intermediate states) respectively (see text for details). All masses are in GeV.}
\label{tab:bp}
\end{table}

We have generated the SUSY spectrum using the publicly available
code \texttt{SuSpect}\cite{Djouadi:2002ze}.  The branching ratios of the relevant
sparticles have been computed using \texttt{SUSYHIT}\cite{Djouadi:2006bz}.  Since
three-body decay modes of the left-sleptons are not computed by
\texttt{SUSYHIT}, we have used \texttt{calcHEP}\cite{Belyaev:2012qa} to compute them. 
We define SS3L+ X as our signal, where X does not include 
$l$ or $\bar{l}$. The rates for this signal are calculated for both the 13 and 14 TeV
runs of the LHC.

We have used \texttt{Prospino}\cite{Beenakker:1996ed} to obtain the NLO cross-sections 
for $\tilde{t}_1 \tilde{t}_1^*$ and $\tilde{g} \tilde{g}$ production at the
LHC. \texttt{MADGRAPH}\cite{Alwall:2011uj}has been used for event generations; subsequent
decays, showering and hadronization has been taken care of by
\texttt{PYTHIA}\cite{Sjostrand:2006za}; \texttt{FASTJET}\cite{Cacciari:2011ma} and 
\texttt{DELPHES}\cite{deFavereau:2013fsa} has been used
for jet clustering (using anti-$k_T$ algorithm) and (ATLAS) detector
simulation respectively. We have used \texttt{MADANALYSIS} \cite{Conte:2012fm} 
to analyse the events.

The signal event selection criteria are: 
\begin{enumerate}
\item $E^j_T > 20$ GeV; $|\eta_j|,  |\eta_l| < 2.5$.
 \item Lepton-lepton separation $\Delta R_{ll} > 0.2$;  lepton-jet
   separation $\Delta R_{lj} > 0.4$, where $\Delta R = \sqrt{\Delta
     \eta^2 + \Delta \phi^2}$.
 \item   For leptons in decreasing order of hardness, $p_T \ge 30, 30, 15$ GeV;
  \item Missing transverse energy $MET > 100$ GeV;
  \item $E_T^{hadron}/E_T^{lepton} \leq 0.1$  within a cone of $\Delta R \leq 0.2$
    around each electron; $\Sigma E_T^{hadron} \leq 1.8$ GeV within a similar
    cone around each muon.
  \item The electron and muon detection efficiencies are
  taken as 85\% - 95\% (following DELPHES). 
\end{enumerate}

The background for SS3L from the Standard Model is negligibly small.  
It has been computed using \texttt{ALPGEN}\cite{Mangano:2002ea} with similar cuts 
mentioned above\cite{Mukhopadhyaya:2010qf}.  The standard model cross-section for 
SS3L events is $\lsim 2.5 \times 10^{-3}$ fb, to which $t \bar{t}W$ contributes 
the most. However, some background may come from standard
model processes with (a) lepton charge misidentification, and (b) jets
faking as leptons. Imposing the $MET$ cut of 100 GeV, which generically
reduces standard model contributions, the total background to SS3L is indeed
negligible. Note that, for the kind of LSP masses considered, one can in principle
raise the MET cut even higher without really affecting the signal, and thus the 
backgrounds can threaten us even less.

\begin{table}[h]
\small
\begin{tabular}{|c|c|c|c|c|}
\hline BP & \multicolumn{2}{c|}{13 TeV} & \multicolumn{2}{c|}{14 TeV}
\\ \hline \hline & $\tilde{t}_{1}\tilde{t}^{*}_{1}$ &
$\tilde{g}\tilde{g}$ &
\multicolumn{1}{c|}{$\tilde{t}_{1}\tilde{t}^{*}_{1}$} &
$\tilde{g}\tilde{g}$ \\ \hline \hline $A$
& \begin{tabular}[c]{@{}c@{}}$5.8 \pm 3.4$\\ ($51.66 \pm
    9.94$)\end{tabular} & \begin{tabular}[c]{@{}c@{}}$13.88 \pm
  5.24$\\ ($60.16 \pm 10.70$)\end{tabular}
& \begin{tabular}[c]{@{}c@{}}$8.38 \pm 4.08$\\ ($67.24 \pm
    11.36$)\end{tabular}& \begin{tabular}[c]{@{}c@{}}$22.50 \pm
  6.66$\\ ($94.04 \pm 13.4$)\end{tabular} \\ \hline \hline

$B$ & \begin{tabular}[c]{@{}c@{}}$4.44 \pm 2.98$\\ ($43.84 \pm
  9.18$)\end{tabular} & \begin{tabular}[c]{@{}c@{}}$13.4 \pm
  5.14$\\ ($54.64 \pm 10.22$)\end{tabular}
& \begin{tabular}[c]{@{}c@{}}$7.90 \pm 3.96$\\ ($62.30 \pm
    10.94$)\end{tabular}& \begin{tabular}[c]{@{}c@{}}$18.08 \pm
  5.98$\\ ($89.58 \pm 13.08$)\end{tabular} \\ \hline \hline

$C$ & \begin{tabular}[c]{@{}c@{}}$3.62 \pm 2.68$\\ ($34.60 \pm
  8.20$)\end{tabular} & \begin{tabular}[c]{@{}c@{}}$8.38 \pm
  4.08$\\ ($52.04 \pm 9.98$)\end{tabular}
& \begin{tabular}[c]{@{}c@{}}$2.96 \pm 2.44$\\ ($50.48 \pm
    9.90$)\end{tabular}& \begin{tabular}[c]{@{}c@{}}$16.02 \pm
  5.64$\\ ($85.12 \pm 12.76$)\end{tabular} \\ \hline \hline

\end{tabular}
\caption{Estimated number of SS3L (SS2L) events for 13 and 14 TeV LHC
  (with 100 $fb^{-1}$ of integrated luminosity) from cascade decays of
  $\tilde{t}_1 \tilde{t}_1^*$ and $\tilde{g} \tilde{g}$ after applying
  the relevant cuts. Note that for SS2L events both leptons are
  required to have $p_T > 30$ TeV.}
\label{tab:res}
\end{table}

In Table \ref{tab:res}, we list the number of SS3L events
with an integrated luminosity of 100 $fb^{-1}$, for both the 13 and 
14 TeV runs, for the three benchmark points chosen above.
Contributions from both $\tilde{t}_1 \tilde{t}_1^*$ direct production
and gluino-pairs are shown separately.  The total number
of SS3L events can be estimated by
adding the contributions from each of these initial states
($\tilde{g}\tilde{g}$ and $\tilde{t}_1 \tilde{t}_1^*$) together.
The corresponding number of same-sign dilepton (SS2L) events
are also shown within parenthesis. The corresponding background
at 14 TeV can be brought under control with a MET cut of 100 GeV,
and an appropriate hardness cut ($\gsim$ 30 GeV),
as used in our analysis \cite{Baer:1995va,Berger:2013sir}. Clearly, 
while direct stop-pair production channel is sufficient to yield background-free SS3L events
that can be detected with the integrated luminosity of one-or
two hundred $fb^{-1}$, the rate goes up several times through
gluino pair-production. This is due to (i) the colour and spin multiplicity
of the gluino, and (ii) the Majorana nature of the gluino, 
which yields leptons of either sign with equal probability in the cascade.
It should also be noticed that BP A, B and C have progressively decreasing
SS3L rates, the reason for which has been explained earlier. While the
presence of a light chargino in BP B causes  the loss of some events,
the loss is more in BP C where the light stop has more left chiral component.
On the whole, however, one obtains distinguishable SS3L rates, even for
a gluino as massive as 1.6 TeV. Relatively heavier stops (which are lighter
than the gluino) do not affect the total number of events very significantly. 
It should also be noted that the rate of SS3L events drop drastically 
if the positions of the neutralinos (at least two) and a chargino are swapped 
with the left-slepton doublet in the spectrum. This is because the left-sleptons are 
produced much more restrictively from strong sparticle production processes.

\begin{figure}[ht!]
\begin{center}
\includegraphics[scale=0.35,angle=-0]{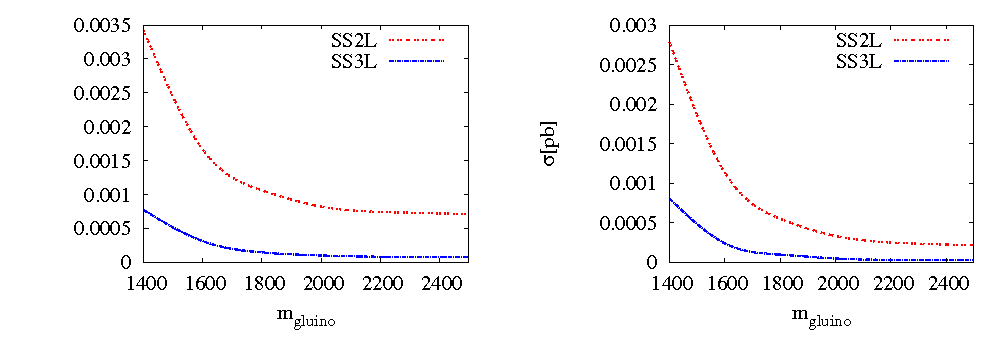}
\end{center}
~~~~~(a)~~~~~~~~~~~~~~~~~~~~~~~~~~~~~~~~~~~~~~~~~~~(b)
\caption{In the left panel the effective NLO cross-sections for SS3L and
  SS2L events (for LHC 14 TeV run) have been plotted against the
  gluino mass assuming the simplest mass hierarchy as shown in figure
  \ref{fig:spec}. (a) shows the relevant numbers for
  $m_{\tilde{t}_1} =1000$ GeV, while  (b) demonstrates the
  same for $m_{\tilde{t}_1} =1200$ GeV. }
\label{fig:res}
\end{figure}

The features mentioned at the end of the last paragraph become obvious
in Figure 1, where the signal rates are plotted against the gluino mass
for two values of the lightest stop mass. The SS3L signal remains
detectable for the gluino mass upto 2 TeV or more, somewhat marginally 
with an integrated luminosity of 100 TeV but rather strongly with twice that
luminosity. 

It may be contended that SS2L events, which are obviously more copious, render
the SS3L events redundant. However, it should be borne in mind that the point
under investigation here is the discernibility of a sneutrino dark matter scenario.
Since this scenario in its common form is all but ruled out, its observation is a rather
striking phenomenon which has wide implication in dark matter 
physics. The predicted SS3L signal makes this new scenario
testable at the LHC.

\begin{table}[h]
\begin{tabular}{|c|c|c|}
\hline
BP & \multicolumn{2}{c|}{0l + 2j}                                                           \\ \hline \hline
   & $\tilde{t}_{1}\tilde{t}^{*}_{1}$ & $\tilde{g}\tilde{g}$                               \\ \hline
   $A$  & $272.7 \pm 13.5$ & $125.3 \pm 10.5$
\\ \hline
$\chi^0_1$-LSP case  & $460.6 \pm 14.2$    & $422.2 \pm 15.5 $ \\
\hline
   
\end{tabular}
\caption{Estimated number of $0l+ \geq 2j$ events at NLO for 14 TeV LHC 
(with 100 $fb^{-1}$ of integrated luminosity) from cascade decays of $\tilde{t}_1 \tilde{t}_1^*$ 
and $\tilde{g} \tilde{g}$ after applying the relevant cuts. Benchmark B represents the same 
scenario as benchmark A with the first two generations of slepton doublets decoupled. Thus 
$\chi_1^0$ is the LSP in benchmark B.}
\label{tab:res1}
\end{table}

Another new feature of this scenario is demonstrated in Table
III, where we present the rates for zero-lepton events (with MET) for
BP A. The numbers of events, corresponding to both the $\tilde{t}_1
\tilde{t}_1^*$ and $\tilde{g} \tilde{g}$ channels, are compared with
the corresponding case with $\chi_1^0$-LSP, where the charged leptons
as well as sneutrinos are decoupled. It is clear from the table that 
the number of hadronic events becomes less than half in the situation with
a sneutrino LSP, and thus the search limit based on such events are
lowered in this case.

The same conclusion also holds when one goes beyond the MSSM spectrum, and 
there is a lighter axino or a gravitino. The sneutrino decays invisibly in that 
case, and all the results presented above are equally valid. Thus 
SS3L also constitutes the most distinct signal of the axino/gravitino LSP, 
sneutrino NLSP scenario. As mentioned above, similar scenarios have been 
studied \cite{Katz:2009qx,Katz:2010xg,Kribs:2008hq} earlier. Of these, SS3L 
has been mentioned in \cite{Katz:2010xg} when the gravitino or the axino has 
to be necessarily present there, as well in \cite{Katz:2009qx,Kribs:2008hq}. 
Moreover, it may be possible to obtain SS3L events without necessarily having 
a light slepton doublet, for example, if $\tilde{t}_1 \rightarrow t~\chi_i^0$ and 
$\chi_i^0 \rightarrow \chi_1^\pm W^\mp$, in certain possibly tuned MSSM scenarios. 
Since in such cases leptons are produced from $W$ bosons 
(on or off-shell), the resulting SS3L events will be flavor blind. On the other hand, 
the presence of a light slepton doublet, as in the present context, would assure 
an excess of SS3L events or SS2L events with the leptons sharing the flavor 
of the light slepton doublets. Of-course, if all three generations of sleptons are 
light (and degenerate) then such a distinctive feature will be absent.
The current study will hopefully bring out the full implication of SS3L in the 
context set here.

To conclude, we have considered {\it an MSSM spectrum with a 
$\tilde{\nu}_L$ dark matter}. The viability of this is assured with,
for example, a split between the scalar and pseudoscalar parts of
$\tilde{\nu}_L$, thus opening up a distinct SUSY dark matter scenario,
finding whose experimental signature is crucial.Thanks to the close
proximity between $\tilde{l}_L$ and $\tilde{\nu}_L$ states demanded by
SU(2), SUSY cascades can lead to SS3L events, via decays of the top
quark or a chargino. At the same time. the $jets + 0\ell + MET$ signal
suffers from suppression, since SUSY cascades leading to $\chi_1^0$
end up in charged sleptons and leptons in a significant fraction of
cases. Thus the importance of leptonic SUSY signals increases, and,
among them, the SS3L events serve as a useful diagnostic. 
 
We estimate the number of such events at the 13 and 14 TeV runs of the
LHC, and show that they can are detectable for gluino masses exceeding 2 TeV,
for integrated luminosities around 100 $fb^{-1}$ or a little higher up. A detailed study
of the SUSY parameter space in such a scenario, including all signals with and
without isolated leptons, will be presented in a later work.

{\bf Acknowledgment:} We would like to thank J. Beuria,
A. Choudhury and N. Sahu for helpful discussions.  This work
was partially supported by funding available from the Department of
Atomic Energy, Government of India for the Regional Centre for
Accelerator-based Particle Physics, Harish-Chandra Research Institute.



%

\bibliography{ref.bib}
\end{document}